\documentclass[a4paper]{article}

\usepackage[english]{babel}
\usepackage[square,numbers]{natbib}
\usepackage[utf8x]{inputenc}
\usepackage[T1]{fontenc}
\usepackage{float}
\usepackage{scalerel}
\usepackage[a4paper,top=2cm,bottom=2cm,left=3cm,right=3cm,marginparwidth=2cm]{geometry}
\usepackage{ragged2e}
\usepackage{amsmath}
\usepackage{authblk}
\usepackage{graphicx}
\usepackage{url}
\usepackage[colorinlistoftodos]{todonotes}
\usepackage[colorlinks=true, allcolors=blue]{hyperref}

\numberwithin{equation}{subsection}

\title{\textbf{Non-Relativistic Scenario of a Free Particle in an Infinite Cylindrical Potential Well}}
\author{Pratik Adarsh}
\author{Sabyasachi Ghosh}
\affil{Department of Physics, Indian Institute of Technology Bhilai, GEC Campus, Sejbahar, Raipur, Chhattisgarh 492015, India}
\date{}
\begin{document}	
\maketitle
\justify{
\begin{abstract}
There are various types of infinite potential well problems occurring in elementary quantum mechanics formalism. The infinite square well (one dimensional), cubical box and, spherical well are quite common in textbooks. In this paper, we consider a rather uncommon potential well, infinite cylindrical well, and try to find its energy eigenvalues and eigenfunctions using Schr\"{o}dinger equation. These results are also derived in the paper\cite{Baltenkov_2016}. But our approach here is to elaborate every step just the way Griffiths did in his book\cite{griffiths2010introduction} for hydrogen atom, using spherical potential well. We also plot some radial wavefunctions and density plots.
\end{abstract}

\section{Introduction}\label{intro}
In infinite spherical potential well, we readily solve the time-independent Schr\"{o}dinger equation to get energy eigenvalues and eigenfunctions. In this problem of infinite cylindrical well, we proceed with the same approach by solving the time-independent Schr\"{o}dinger equation in cylindrical coordinates.\\
The motivation is to see how cylindrical symmetry differs from spherical one and to what extend its analytical solutions are possible in the most general way.

\section{The Problem}\label{prob}

\begin{figure}[H]
\centering
\includegraphics[width=0.4\textwidth]{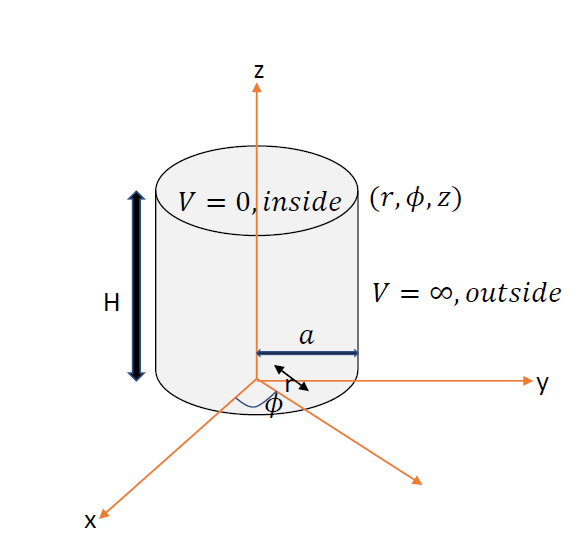}
\caption{\label{fig:fig1}Infinite cylindrical well.}
\end{figure}


We have a particle inside a cylindrical-shaped well for which the potential is zero inside the boundaries but infinite at and outside the boundaries.\vspace*{0.1 in}
The dimensional attributes of the well are:\\
Height of the cylinder = $H$\\
Radius of the cylinder = $a$\\
With these we have to solve the time-independent Schr\"{o}dinger equation:\\
$${\frac{-{\hbar}^2}{2m}}{{\nabla}^2}{\psi}+{V\psi}=E\psi$$
\section{The Approach}\label{aprch}

\subsection{Separation of Variables}\label{sub3.1}
The Laplacian in cylindrical polar coordinates is:\\
\begin{equation*}
{\nabla^2}= {\frac{1}{r}}{\frac{\partial}{\partial r}}\left({r}\frac{\partial}{\partial r}\right)+{\frac{1}{r^2}}{\frac{\partial^2}{\partial \phi^2}}+\frac{\partial^2}{\partial z^2}
\end{equation*}
The wavefunction is zero outside the well, so the Schr\"{o}dinger equation inside($V=0$) the well reads:\\
\begin{equation}\label{311}
\frac{-\hbar^2}{2m}\left[\frac{1}{r}\frac{\partial}{\partial r}\left(r\frac{\partial\psi}{\partial r}\right)+\frac{1}{r^2}\frac{\partial^2}{\partial \phi^2}+\frac{\partial^2 \psi}{\partial z^2}\right]=E\psi
\end{equation}
We let the solutions separable into products:
\begin{equation*}
\psi(r,\phi,z)=\psi_{r \phi}(r,\phi)\psi_z(z)
\end{equation*}
Substituting this into equation \ref{311},we have
\begin{equation}\label{312}
\frac{-\hbar^{2}}{2 m}\left[\frac{\psi_{z}}{r} \frac{\partial}{\partial r}\left(r \frac{\partial \psi_{r \phi}}{\partial r}\right)+\frac{\psi_{z}}{r^{2}} \frac{\partial^{2} \psi_{r \phi}}{\partial \phi^{2}}+\psi_{r \phi} \frac{\partial^{2} \psi_{z}}{\partial z^{2}}\right]=E\psi_{r \phi}\psi_z
\end{equation}
Rearranging terms, we get
\begin{equation}\label{313}
\underbrace{\frac{1}{\psi_{r \phi}}\left[\frac{1}{r} \frac{\partial \psi_{r \phi}}{\partial r}+\frac{\partial^{2} \Psi_{r \phi}}{\partial r^{2}}+\frac{1}{r^{2}} \frac{\partial^{2} \psi_{r \phi}}{\partial \phi^{2}}\right]}_{\text{depends on }r\text{ and }\phi}+\underbrace{\frac{1}{\psi_{z}}\left[\frac{d^{2} \psi_{z}}{d z^{2}}\right]}_{\text{depends on }z}=-\frac{2mE}{\hbar^2}
\end{equation}
In the L.H.S of equation \ref{313}, the first term depends only on $r$ and $\phi$. Whereas the second term depends only on $z$. So we let the first term equals some factor $A_1$(say)and the second term equals to some another factor $A_2$. With these, the equation \ref{313} reads as:\\
\begin{equation}\label{314}
A_1+A_2=\frac{-2mE}{\hbar^2}
\end{equation}
This equation \ref{314} is the main equation that would give us the energy eigenvalues for the system. Our aim would be now to find these factors, $A_1$ and $A_2$.\\
Coming back to equation \ref{313}, we write the first term on the L.H.S as:
\begin{equation}\label{315}
\frac{1}{r} \frac{\partial \psi_{r \phi}}{\partial r}+\frac{\partial^{2} \Psi_{r \phi}}{\partial r^{2}}+\frac{1}{r^{2}} \frac{\partial^{2} \psi_{r \phi}}{\partial \phi^{2}}=A_1\psi_{r \phi}
\end{equation}
This is just another partial differential equation we have to solve for proceeding in the right direction. We let the solutions separable like previously:\\
\begin{equation*}
\psi_{r \phi}(r,\phi)=R(r)X(\phi)
\end{equation*}
Substituting this into equation \ref{315}, we have
\begin{equation}\label{316}
\frac{X}{r} \frac{d R}{d r}+X \frac{d^{2} R}{d r^{2}}+\frac{R}{r^{2}} \frac{d^{2} X}{d \phi^{2}}=A_{1} R X
\end{equation}
Rearranging terms, we get
\begin{equation}\label{317}
\underbrace{\left[\frac{r}{R} \frac{d R}{d r}+\frac{r^2}{R} \frac{d^{2} R}{d r^{2}}-A_1r^2\right]}_{\text{depends on }r}+\underbrace{\left[\frac{1}{X} \frac{d^{2} X}{d \phi^{2}}\right]}_{\text{depends on }\phi}=0
\end{equation}
The first term on L.H.S of the equation \ref{317} depends only on $r$ and the second term depends only on $\phi$. So we let the first term equals to some factor $B$(say), then obviously the second term equals to $-B$.\\
With these attributes, we have:\\
\begin{equation}\label{318}
\frac{r}{R} \frac{d R}{d r}+\frac{r^2}{R} \frac{d^{2} R}{d r^{2}}-A_1r^2=B
\end{equation}
\begin{equation}\label{319}
\frac{1}{X} \frac{d^{2} X}{d \phi^{2}}=-B
\end{equation}
And from equations \ref{313} and \ref{314}
\begin{equation}\label{3110}
\frac{1}{\psi_{z}}\left[\frac{d^{2} \psi_{z}}{d z^{2}}\right]=A_2
\end{equation}
\subsection{The Radial Equation}\label{radeq}
We can rewrite equation \ref{318} as:
\begin{equation}\label{321}
\frac{d}{dr}\left(rR'(r)\right)+\left[-A_1-\frac{B}{r^2}\right]rR(r)=0 \textit{ (Strum-Liouville Equation)}
\end{equation}
Its solution is:
\begin{equation}\label{322}
R(r)=C_1J_{\scaleto{\sqrt{B}}{5pt}}\left(-i\sqrt{A_1}r\right)+C_2Y_{\scaleto{\sqrt{B}}{5pt}}\left(-i\sqrt{A_1}r\right)
\end{equation}
Where $J$ and $Y$ are Bessel functions of first and second kind respectively. Equation \ref{322} gives us the radial wavefunction. We have to solve it with appropriate boundary conditions. We would do this later, now the main task is to simplify it with this simple view:\\
For $r\,\to\,0$, equation \ref{322} reads as:\\
\begin{equation*}
R(0)=C_1J_{\scaleto{\sqrt{B}}{5pt}}\left(0\right)+C_2Y_{\scaleto{\sqrt{B}}{5pt}}\left(0\right)
\end{equation*}
But we know that
\begin{equation*}
Y_{\scaleto{\sqrt{B}}{5pt}}\left(0\right)\,\to\,\infty
\end{equation*}
\begin{equation*}
J_{\scaleto{\sqrt{B}}{5pt}}\left(0\right)=0
\end{equation*}
This means that the radial wavefunction, $R(0)$ diverges on the axis of the well. In other words this implies that it's a non-normalizable wavefunction.\\
Therefore we must have $C_2=0$.\\
So now our radial wavefunction, equation \ref{322} becomes:
\begin{equation}\label{323}
R(r)=C_1J_{\scaleto{\sqrt{B}}{5pt}}\left(-i\sqrt{A_1}r\right)
\end{equation}
We will return to this equation soon for calculating $A_1$, to plug in the equation \ref{314} and getting the energy eigenvalues. But first, we have to find what $B$ corresponds and of course, we also have to find $A_2$.
\subsection{The Angular Equation}\label{angeq}
The equation \ref{319} is:\\
\begin{equation*}
\frac{1}{X} \frac{d^{2} X}{d \phi^{2}}=-B
\end{equation*}
Its solution can be written as:
\begin{equation}\label{331}
X(\phi)=C_3\exp\left(i\sqrt{B}\phi\right)+C_4\exp\left(-i\sqrt{B}\phi\right)
\end{equation}
For rotational symmetry, we must have:
\begin{equation*}
X(\phi+2\pi)=X(\phi)
\end{equation*}
\begin{equation*}
\implies C_3\exp\left(i\sqrt{B}(\phi+2\pi)\right)+C_4\exp\left(-i\sqrt{B}(\phi+2\pi)\right)=C_3\exp\left(i\sqrt{B}\phi\right)+C_4\exp\left(-i\sqrt{B}\phi\right)
\end{equation*}
Simplifying and equating real and imaginary parts, we get:\\
\begin{equation*}
\left(C_3+C_4\right)\left[\cos(\sqrt{B}\phi+2\pi\sqrt{B})\right]=\left(C_3+C_4\right)\cos(\sqrt{B}\phi)
\end{equation*}
and
\begin{equation*}
\left(C_3-C_4\right)\left[\sin(\sqrt{B}\phi+2\pi\sqrt{B})\right]=\left(C_3-C_4\right)\sin(\sqrt{B}\phi)
\end{equation*}
Therefore $\sqrt{B}$ is an integer, for $C_3+C_4 \neq 0$ and $C_3 \neq C_4$. Hence we have
\begin{equation}\label{332}
\sqrt{B}=n_\phi=0,1,2,...
\end{equation}
where $n_\phi$ denotes a quantum number associated with azimuthal angle.
\subsection{The Equation along Z Direction}\label{zeq}
The equation \ref{3110} is:
\begin{equation*}
\frac{1}{\psi_{z}}\left[\frac{d^{2} \psi_{z}}{d z^{2}}\right]=A_2
\end{equation*}
Its solution can be written as:
\begin{equation}\label{341}
\psi_z(z)=C_5\exp\left(iz\sqrt{\frac{2mE}{\hbar^2}+A_1}\right)+C_6\exp\left(-iz\sqrt{\frac{2mE}{\hbar^2}+A_1}\right)
\end{equation}
Now we have the following boundary conditions:
\begin{enumerate}
\item[(\textit{i.})]$\psi_z(0)=0$
\item[(\textit{ii.})]$\psi_z(H)=0$
\end{enumerate}
Imposing the first condition on the equation \ref{341}, we have:\\
\begin{equation*}
C_6=-C_5
\end{equation*}
With this and by second boundary condition, the equation \ref{341} becomes:
\begin{equation*}
C_5\exp\left(iz\sqrt{\frac{2mE}{\hbar^2}+A_1}\right)-C_5\exp\left(-iz\sqrt{\frac{2mE}{\hbar^2}+A_1}\right)=0
\end{equation*}
\begin{equation*}
\implies 2iC_5\left[\sin\left(H\sqrt{\frac{2mE}{\hbar^2}+A_1}\right)\right]=0
\end{equation*}
\begin{equation}\label{342}
\implies H\sqrt{\frac{2mE}{\hbar^2}+A_1}=n_z\pi\hspace{5pt}; n_z=1,2,3,...
\end{equation}
Where $n_z$ denotes a quantum number associated with z-direction.\\
Using equations \ref{342} and \ref{314}, we have:
\begin{equation}\label{343}
A_2=\frac{-n_z^2\pi^2}{H^2}
\end{equation}
Therefore we have energy expression redefined as:
\begin{equation}\label{344}
E=\frac{\hbar^2(n_z^2\pi^2-A_1H^2)}{2mH^2}
\end{equation}
\subsection{Energy Eigenvalues}\label{energyev}
With some remaining calculations, we are now very near to get the energy eigenvalues.\\
It's now time to return to equation \ref{323} with the consideration of equation \ref{332}, we have
\begin{equation}\label{351}
R(r)=C_1J_{\scaleto{n_\phi}{5pt}}\left(-i\sqrt{A_1}r\right)
\end{equation}
Imposing the boundary condition\\
\begin{equation*}
R(a)=0
\end{equation*}
We get\\
\begin{equation*}
\implies R(a)=C_1J_{\scaleto{n_\phi}{5pt}}\left(-i\sqrt{A_1}a\right)=0
\end{equation*}
\begin{equation}\label{352}
\implies A_1=\frac{-(j_{\scaleto{n_\phi,n_r}{5pt}})^2}{a^2}\hspace{5pt}; n_r=1,2,3,...
\end{equation}
Where $n_r$ denotes a quantum number associated with radial solution and $j_{\scaleto{n_\phi,n_r}{5pt}}$ denotes $n_r^{th}$ root of Bessel function $J_{\scaleto{n_\phi}{5pt}}$\\
Putting this equation \ref{352} into equation \ref{344}, we have the final expression for energy eigenvalues as:

\begin{equation}\label{353}
\boxed{E_{\scaleto{n_r,n_\phi,n_z}{5pt}}=\frac{\hbar^2\left[n_z^2a^2\pi^2+(j_{\scaleto{n_\phi,n_r}{5pt}})^2H^2\right]}{2ma^2H^2}}
\end{equation}
Where
\begin{align*}
n_r &= 1,2,3,...\\
n_\phi &= 0,1,2,...\\
n_z &= 1,2,3,...
\end{align*}
\subsection{The Eigenfunctions}\label{efunc}
We started with separable solutions of the Schr\"{o}dinger's equation \ref{311} as:
\begin{equation}\label{361}
\psi(r,\phi,z)=\psi_{r \phi}(r,\phi)\psi_z(z)=R(r)X(\phi)\psi_z(z)
\end{equation}
Now we will find $R(r)$, $X(\phi)$ and $\psi_z(z)$ separately and put them back into the equation \ref{361} to get the complete eigenfunctions.\\

\textbf{\textit{The Normalization Technique}}\\

Before finding the wavefunctions, we emphasize that how normalization can be done separately for each of them.\\
In cylindrical coordinates, we have elementary volume as:
\begin{equation*}
dV=rdr d\phi dz
\end{equation*}
So normalization of wavefunction in the view of equation \ref{361} becomes
\begin{equation*}
\iiint_V \psi(r,\phi,z)\psi^*(r,\phi,z){r dr d\phi dz}=\int_{0}^{a}R(r)R^*(r)r dr\int_{0}^{2\pi}X(\phi)X^*(\phi)d\phi\int_{0}^{H}\psi_z(z)\psi_z^*(z)dz=1
\end{equation*}
We can normalize each wavefunctions separately as follows:
\begin{enumerate}
\item[(\textit{i.})]\begin{equation}\label{362}\int_{0}^{a}R(r)R^*(r)r dr=1
\end{equation}
\end{enumerate}
\begin{enumerate}
\item[(\textit{ii.})]\begin{equation}\label{363}
\int_{0}^{2\pi}X(\phi)X^*(\phi)d\phi=1
\end{equation}
\end{enumerate}
\begin{enumerate}
\item[(\textit{iii.})]\begin{equation}\label{364}
\int_{0}^{H}\psi_z(z)\psi_z^*(z)dz=1
\end{equation}
\end{enumerate}
\vspace{7pt}
\textbf{\textit{The Radial wavefunctions}}\\

From the equations \ref{351} and \ref{352}, we have:
\begin{equation}\label{365}
R(r)=C_1J_{\scaleto{n_\phi}{5pt}}\left[\frac{(j_{\scaleto{n_\phi,n_r}{5pt}})r}{a}\right]
\end{equation}
To make further calculations easier to follow, we take the following denotations:
\begin{equation*}
n_\phi=n \text{ and } j_{\scaleto{n_\phi,n_r}{5pt}}=\beta
\end{equation*}
With these, the radial equation \ref{365} becomes
\begin{equation}\label{366}
R(r)=C_1J_n\left[\frac{\beta{r}}{a}\right]
\end{equation}
Normalizing using equation \ref{362}, we have
\begin{equation}\label{367}
C_1^2\int_{0}^{a}J_n^2\left[\frac{\beta{r}}{a}\right]r dr=1
\end{equation}
The indefinite integral has solution as:
\begin{equation}\label{368}
\int{J_n^2\left[\frac{\beta{r}}{a}\right]r dr}=\frac{r^2}{2}\left[J_n^2\left[\frac{\beta{r}}{a}\right]-J_{n-1}\left[\frac{\beta{r}}{a}\right]J_{n+1}\left[\frac{\beta{r}}{a}\right]\right]+C
\end{equation}
where $C$ is the constant of integration.\\
With the proper limits of integration, $[0,a]$ and using Bessel recurrence relation:
\begin{equation}\label{369}
J_{n-1}(x)+J_{n+1}(x)=2\frac{n}{x}J_n(x)
\end{equation} 
we have 
\begin{equation*}
C_1=\frac{\sqrt{2}}{aJ_{\scaleto{n+1}{5pt}}\left(\beta\right)}
\end{equation*}
Putting back into equation \ref{365} and restoring the original denotations, we get the normalized radial wavefunctions as:
\begin{equation}\label{3610}
\boxed{R(r)=\frac{\sqrt{2}}{aJ_{\scaleto{n_\phi+1}{5pt}}\left[j_{\scaleto{n_\phi,n_r}{5pt}}\right]}J_{\scaleto{n_\phi}{5pt}}\left[\frac{(j_{\scaleto{n_\phi,n_r}{5pt}})r}{a}\right]}
\end{equation}
\vspace{7pt}
\textbf{\textit{The Angular wavefunctions}}\\

The angular wavefunctions are given by equation \ref{331}, they are basically two linearly independent angular solutions. But we can compress it to the form:
\begin{equation}\label{3611}
X(\phi)=C_p\exp(ip\phi)\text{; where }p=\pm\sqrt{B}=\pm n_\phi=0,\pm1,\pm2,...
\end{equation}
The $p$ is like another quantum number that would be responsible for state degeneracy. We can normalize \ref{3611} in the view of the equation \ref{363}, we get the normalized angular wavefunctions as:
\begin{equation}\label{3612}
\boxed{X(\phi)=\frac{1}{\sqrt{2\pi}}\exp(ip\phi)\text{ ; }p=\pm n_\phi=0,\pm1,\pm2,...}
\end{equation}
\vspace{7pt}
\textbf{\textit{The wavefunctions along Z-Direction}}\\

From subsection \ref{zeq}, we have wavefunctions along z-direction as:
\begin{equation}\label{3613}
\psi_z(z)=2iC_5\left[\sin\left(z\sqrt{\frac{2mE}{\hbar^2}+A_1}\right)\right]
\end{equation}
Using the equation \ref{344}, we can rewrite equation \ref{3613} as:
\begin{equation}\label{3614}
\psi_z(z)=2C_5'\sin\left(\frac{n_z\pi{z}}{H}\right)
\end{equation}
Normalizing in the view of equation \ref{364}, we get the normalized wavefunctions along the z-direction as:
\begin{equation}\label{3615}
\boxed{\psi_z(z)=\left[\sqrt{\frac{2}{H}}\right]\sin\left(\frac{n_z\pi{z}}{H}\right)\text{ ; }n_z=1,2,3,...}
\end{equation}
Now using equation \ref{361}, we stack up all the wavefunctions and get the final expression as:
\begin{equation}\label{3616}
\boxed{\psi_{\scaleto{n_r,n_\phi,n_z,p}{5pt}}=\left[\sqrt{\frac{2}{\pi{Ha^2}}}\right]\frac{1}{J_{\scaleto{n_\phi+1}{5pt}}\left[j_{\scaleto{n_\phi,n_r}{5pt}}\right]}J_{\scaleto{n_\phi}{5pt}}\left[\frac{(j_{\scaleto{n_\phi,n_r}{5pt}})r}{a}\right]\sin\left[\frac{n_z\pi{z}}{H}\right]\exp(ip\phi)}
\end{equation}
Where
\begin{align*}
n_r &= 1,2,3,...\\
n_\phi &= 0,1,2,...\\
n_z &= 1,2,3,...\\
p &= \pm n_\phi
\end{align*}
\section{Plots}\label{plots}
\subsection{Radial Wavefunctions}\label{radplot}
\begin{figure}[H]
\centering
\includegraphics[width=0.8\textwidth]{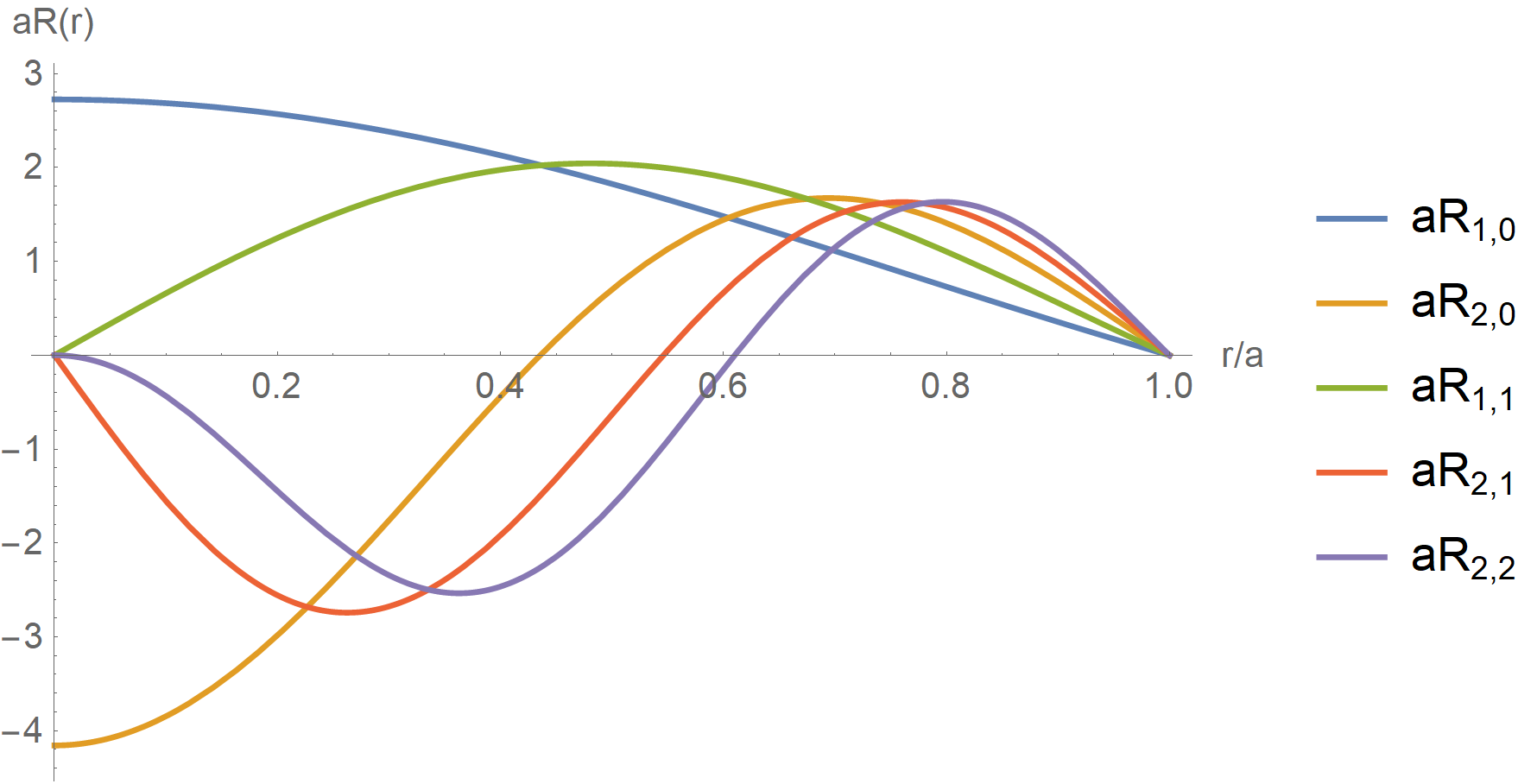}
\caption{\label{fig:fig2}Some Radial Wavefunctions $R_{\scaleto{n_r,n_\phi}{5pt}}$ (\ref{3610}).}
\end{figure}
\subsection{Probability density variation, ${\lvert\psi_{\scaleto{n_r,n_\phi,n_z}{5pt}}\rvert}^2$}\label{probdenvar}
\begin{figure}[!h]
	\centering
	\begin{minipage}{.5\textwidth}
		\centering
		\includegraphics[scale=0.4]{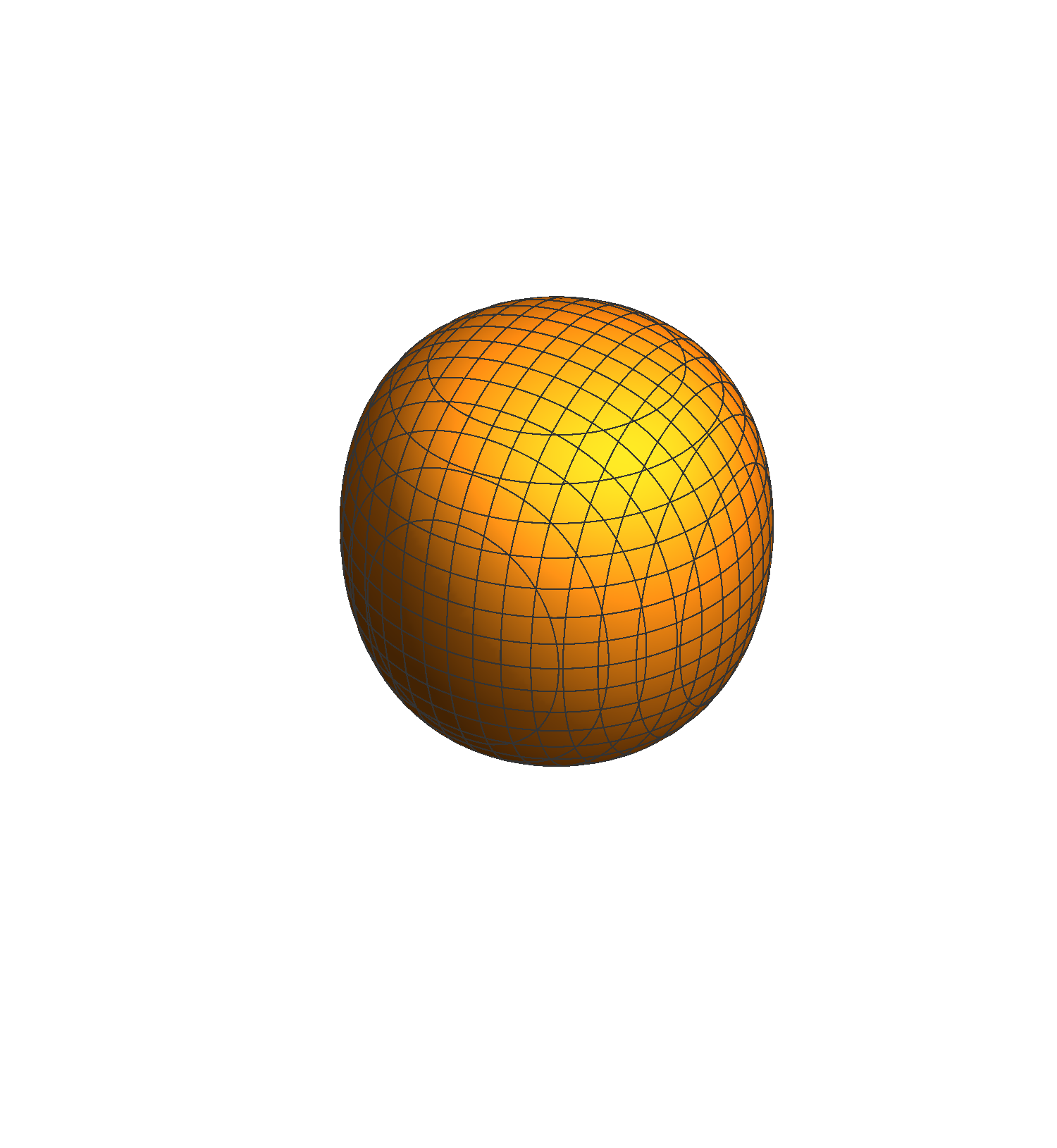}
		\caption{${\lvert\psi_{\scaleto{1,0,1}{5pt}}\rvert}^2$}
	\end{minipage}%
	\begin{minipage}{0.5\textwidth}
		\centering
		\includegraphics[scale=0.4]{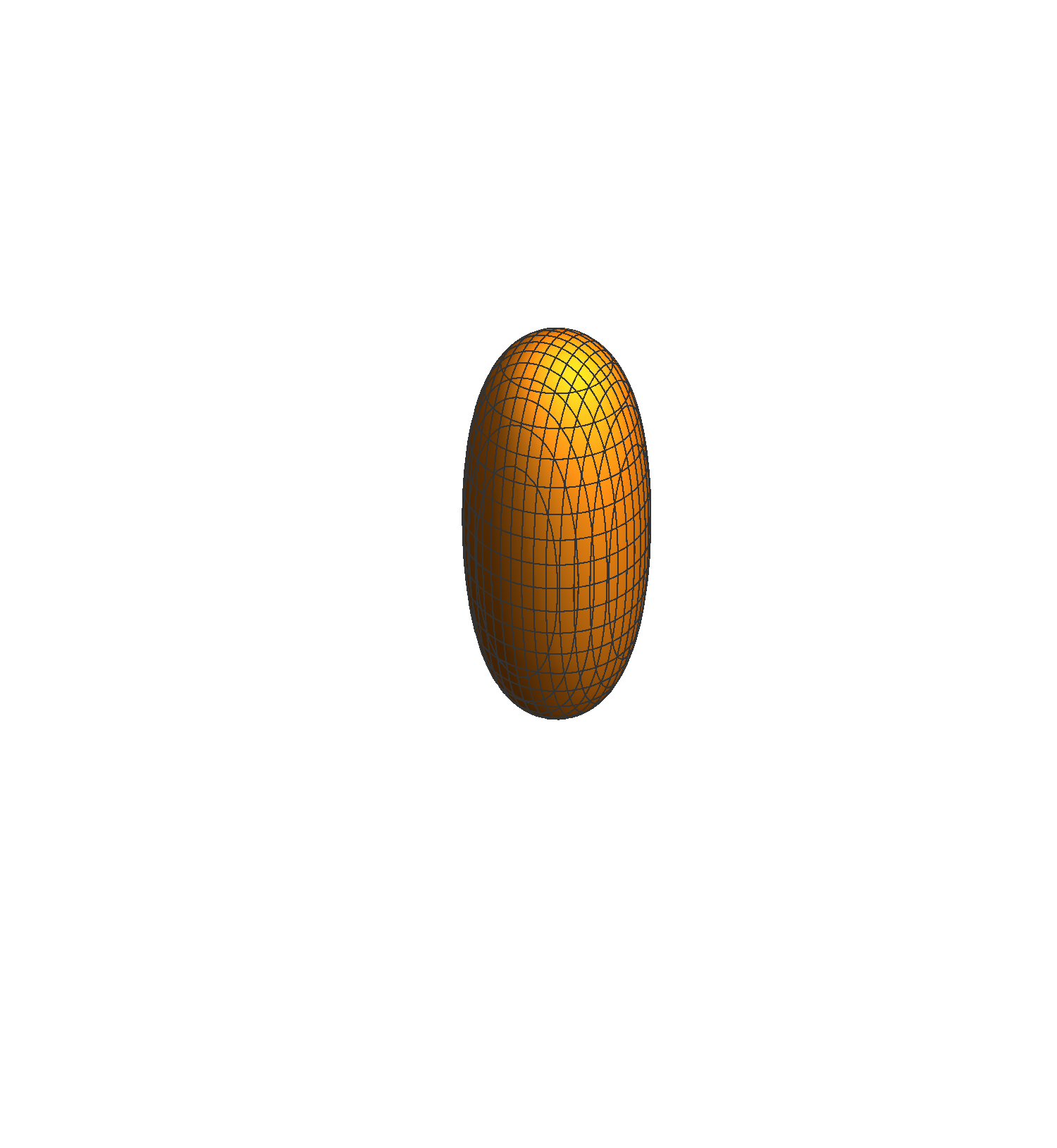}
		\caption{${\lvert\psi_{\scaleto{2,0,1}{5pt}}\rvert}^2$}
	\end{minipage}
\end{figure}

\begin{figure}[!h]
	\centering
	\begin{minipage}{.5\textwidth}
		\centering
		\includegraphics[scale=0.4]{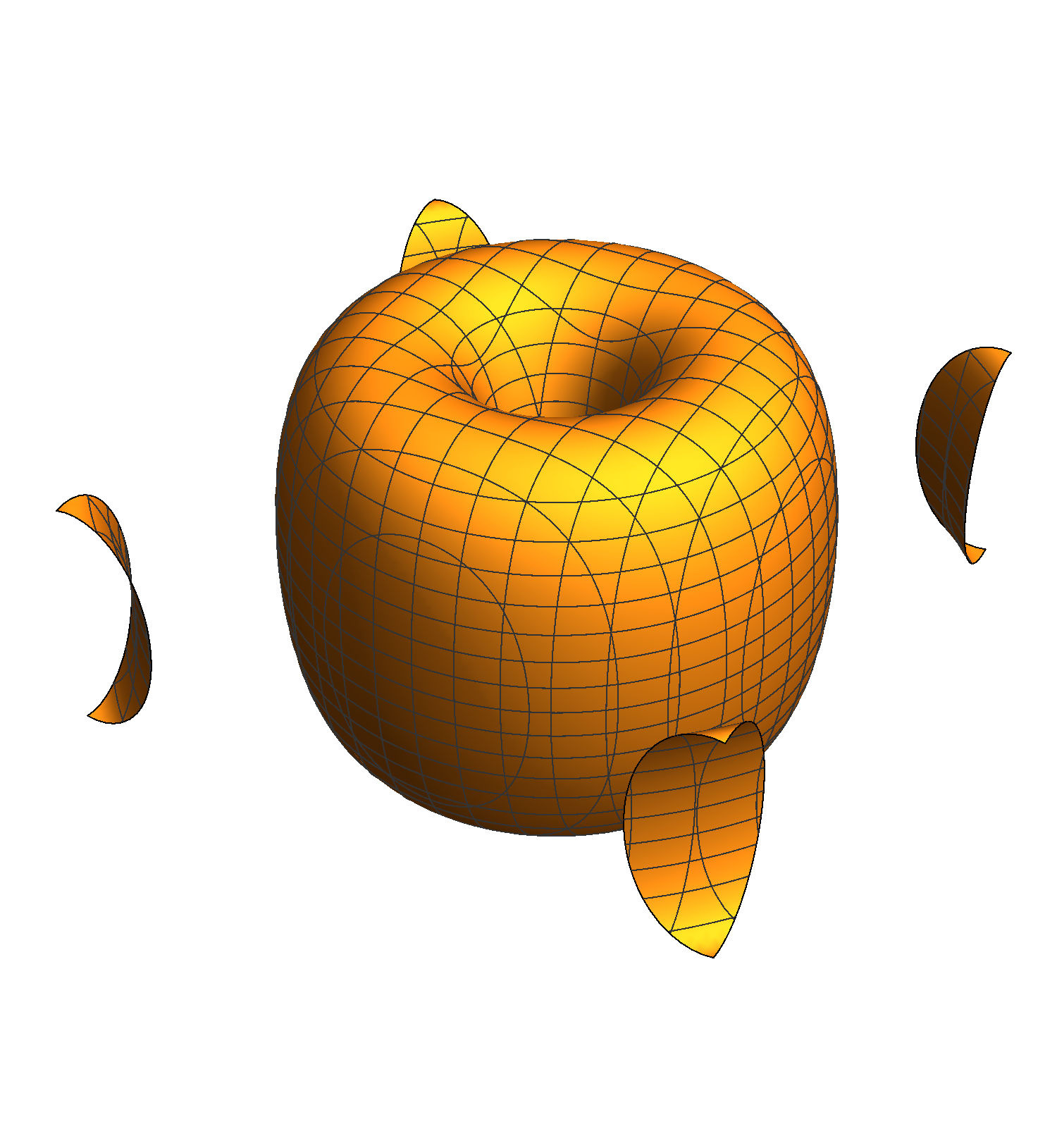}
		\caption{${\lvert\psi_{\scaleto{1,1,1}{5pt}}\rvert}^2$}
	\end{minipage}%
	\begin{minipage}{0.5\textwidth}
		\centering
		\includegraphics[scale=0.4]{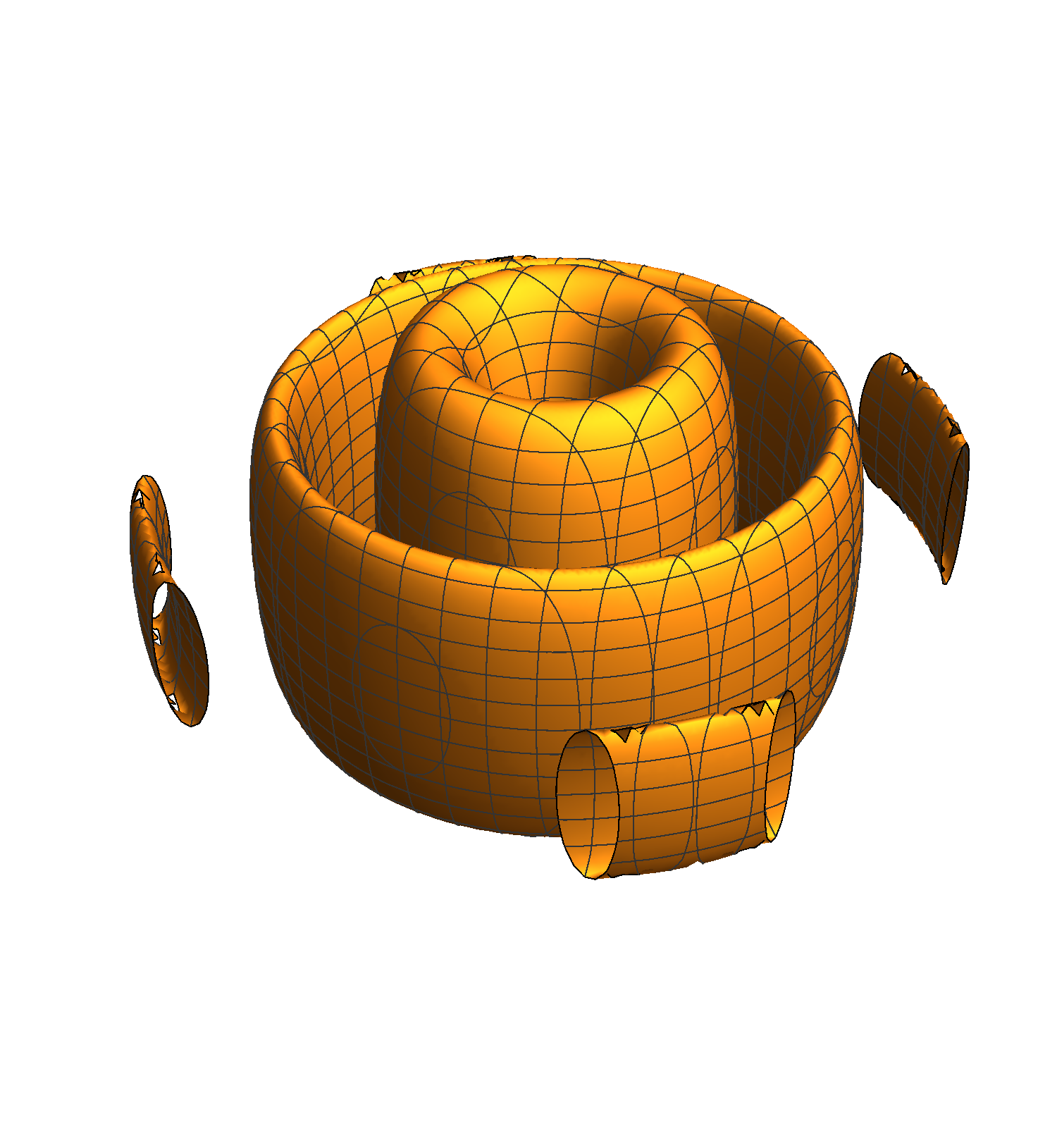}
		\caption{${\lvert\psi_{\scaleto{2,2,2}{5pt}}\rvert}^2$}
	\end{minipage}
\end{figure}
\newpage
\textbf{\textit{Comments}}\\

In \ref{probdenvar} we don't have that extra quantum number $p$, as we had in equation (\ref{3616}). This is obvious as \textit{probability density} $= \psi\psi^* $.\\

In fig. 5 and fig. 6, we have zero probability density on the z-axis. This is evident from the term in (\ref{3616}):
$$J_{\scaleto{n_\phi}{5pt}}\left[\frac{(j_{\scaleto{n_\phi,n_r}{5pt}})r}{a}\right]$$
On the z-axis, $r=0$. So the term becomes
$$J_{\scaleto{n_\phi}{5pt}}\left(0\right)$$
But $$J_n\left(0\right)=0 \text{ if } n\neq0 $$
 $$J_n\left(0\right)=1 \text{ if } n=0 $$
That is how plots in fig. 5 and fig. 6 differ from plots in fig. 3 and fig. 4. These observations can also, be verified with the radial wavefunction plots (\ref{radplot}).\\
In fig. 6, however, we have an additional place (apart from the z-axis) where probability density is vanishing. This can be verified with radial wavefunction plot $R_{\scaleto{n_r,n_\phi}{5pt}}=R_{\scaleto{2,2}{5pt}}$ in (\ref{radplot}).

\section{Conclusions}\label{concl}
We can see from energy eigenvalues expression, equation \ref{353}, that the energy of a free particle in an infinite cylindrical well depends on three quantum numbers associated with three independent directions. But the eigenfunctions, given by equation \ref{3616}, depend on an extra quantum number $p$. For each $n_\phi$, there are two values of $p$. Thus all excited states are 2-fold degenerate.

\bibliographystyle{unsrt}
\bibliography{cite}
	
}
\end{document}